\documentclass[twocolumn,pre]{revtex4-1}

\usepackage{amsmath}
\usepackage{amssymb}
\usepackage{latexsym}
\usepackage{amssymb,amsmath}
\usepackage{float}

\ifx\pdfoutput\undefined
\usepackage{graphicx}
\else
\usepackage[pdftex]{graphicx}
\usepackage{epstopdf}
\fi

\begin{document}
\title{Ecological collapse and the emergence of traveling waves at the onset of shear
turbulence}

\author{Hong-Yan Shih, Tsung-Lin Hsieh, Nigel Goldenfeld \footnote{To whom correspondence should be addressed. E-mail:
nigel@uiuc.edu}
}
\affiliation{Loomis Laboratory of Physics, University of Illinois at Urbana-Champaign, 1110 W.\ Green St.,
 Urbana, IL 61801, USA}

\def\Re{\hbox{Re}}

\newcommand{\etal}{\emph{et al.}{}}




\begin{abstract}
The transition to turbulence exhibits remarkable spatio-temporal
behavior that continues to defy detailed understanding.  Near the onset
to turbulence in pipes, transient turbulent regions decay either
directly or, at higher Reynolds numbers through splitting, with
characteristic time-scales that exhibit a super-exponential dependence
on Reynolds number. Here we report numerical simulations of
transitional pipe flow, showing that a zonal flow emerges at large
scales, activated by anisotropic turbulent fluctuations; in turn, the
zonal flow suppresses the small-scale turbulence leading to stochastic
predator-prey dynamics. We show that this \lq\lq ecological" model of
transitional turbulence reproduces the super-exponential lifetime
statistics and phenomenology of pipe flow experiments.   Our work
demonstrates that a fluid on the edge of turbulence is mathematically
analogous to an ecosystem on the edge of extinction, and provides an
unbroken link between the equations of fluid dynamics and the directed
percolation universality class.

\end{abstract}

\maketitle

\paragraph*{Introduction.}

Fluids in motion are generally found in one of two generic states.  The
most common --- turbulence ---  is found at sufficiently large
characteristic speeds $U$, depending on the kinematic viscosity $\nu$
and the characteristic system scale $D$; turbulent flows are complex,
stochastic, and unpredictable in detail.  At lower velocities, the
fluid is said to be laminar: its flow is simple, deterministic and
predictable.  In between these two states, conventionally delineated by
the dimensionless control parameter known as the Reynolds number
$\Re\equiv UD/\nu$, is a transitional regime that occurs for $1700
\lesssim \Re \lesssim  2300$ in pipes, and which has presented a
challenge to experiment and theory since Osborne Reynolds' original
observation of intermittent \lq\lq flashes" of turbulence
\cite{reynolds}.  Today, Reynolds' flashes are known as puffs
\cite{wygnanski1}, and their behavior has been characterized very
precisely through a series of physical and numerical experiments
performed during the last decade or so
\cite{eckhardt1,peixinho2006decay,hof2006flt,willis1} (for a recent
review, see \cite{song2014deterministic}) culminating in the {\it tour
de force} observation of a super-exponential functional dependence of
the lifetime $\tau^d$ of puffs as a function of Re \cite{hof_lifetime}:
$\ln\ln \tau^d \propto \Re$.  For Reynolds numbers based upon pipe
diameter $D$ of around 2300, turbulence is sustained longer than the
ability to observe its lifetime in finite systems, and the puffs become
unstable through a new dynamical processes in which the leading edge
breaks away and nucleates the formation of a new puff some distance
downstream \cite{wygnanski2,barkley,avila2011onset,nishi}.  The
puff-splitting occurs on a characteristic time that decays
super-exponentially with increasing Re. These phenomena are likely to
be generic.  For example, super-exponential scaling behavior near the
transition to turbulence has also been reported in plane Poiseuille
flow \cite{wesfreid2014} and Taylor-Couette flow
\cite{borrero2009transient}.  In addition, optical analogues of the
laminar-turbulence transition have been observed in laser-driven
optical fibers \cite{falkovich2013}.

The theoretical account of these phenomena has focused primarily on the
existence and interactions between nonlinear solutions of the
Navier-Stokes equations, periodic orbits and streamwise
vortices \cite{willis2013revealing,cvitanovic2013recurrent,avila2013streamwise,KERS05,eckhardt2007,chantry2014genesis},
and the dynamics of long-lived chaotic transients
 \cite{crutchfield1988attractors}.  An alternative line of inquiry has
been to characterize the statistical properties of transitional
turbulence through {\it ad hoc} model equations.  These have been
motivated either by perceptive analogies to excitable media
 \cite{barkley2011simplifying} or by phase transition universality
arguments that begin with the notion that the laminar state is an
absorbing one  \cite{pomeau}, and show quantitatively how
super-exponential decay results from the generic universality class for
non-equilibrium absorbing processes \cite{janssen1981}: directed percolation (DP)
 \cite{sipos2011directed}.  Both approaches reflect an important aspect
of the dynamics, namely that a certain minimum level of energy is
required to sustain turbulent puffs  \cite{peixinho2007finite}, leading
to a further connection with extreme value statistics  \cite{nigel_evs}.

It is the statistical behavior near the transition which concerns us
here: how do the various spatial-temporal modes that are excited give
rise to such remarkable lifetime statistics?  What is the universality
class of this transition, in terms of its fluctuation characteristics?
Are there simplified effective descriptions that bridge the gap between
the underlying fluid dynamics and the large-scale statistical
properties? And how do these emerge from the underlying Navier-Stokes
equations that govern all hydrodynamic phenomena?


The purpose of this report is to address these questions using the same
approach by which phase transitions are understood in condensed matter
physics \cite{goldenfeld1992lectures}.  There it is well-established
that universal aspects of phase transitions, such as the phase diagram,
critical exponents, scaling functions etc. are all described
quantitatively by an effective coarse-grained theory (\lq\lq Landau
theory\rq\rq) that contains the symmetry-allowed collective and
long-wavelength modes, without requiring excessive realism at the
microscopic level of description.  Being based on symmetry principles,
the individual symmetry-allowed terms in Landau theory do not require
detailed derivation from the microscopic level of description.  This is
fortunate given that there is usually no good, uniformly valid
approximation scheme to derive formally and systematically these terms
and their coefficients from first principles.

For the transitional turbulence problem, the analytical difficulties
are acute.  Therefore, in this paper, we avoid approximations which are
difficult to justify systematically.  Instead, we use direct numerical
simulation to identify the important collective modes which exhibit an
interplay between large-scale fluctuations and small-scale dynamics at
the onset of turbulence, and thence to write down the corresponding
minimal stochastic model, in the spirit of the Landau theory of phase
transitions.  As we will show, this coarse-grained description is able
to account for the principal experimental observations and to predict
quantitatively the puff lifetime and splitting behavior observed near
the transition.  Our stochastic model can be transformed using standard
statistical mechanics techniques into a field theory known to be in the
universality class of directed percolation. These results provide an
unbroken link between the equations of fluid dynamics and the directed
percolation universality class \cite{pomeau,sipos2011directed}. Our
approach is thus a precise parallel to the way in which phase
transitions are understood in condensed matter physics, and shows that
concepts of universality and effective theories are applicable to the
laminar-turbulence transition.

\begin{figure}
\begin{center}
\includegraphics[width=0.95\columnwidth]{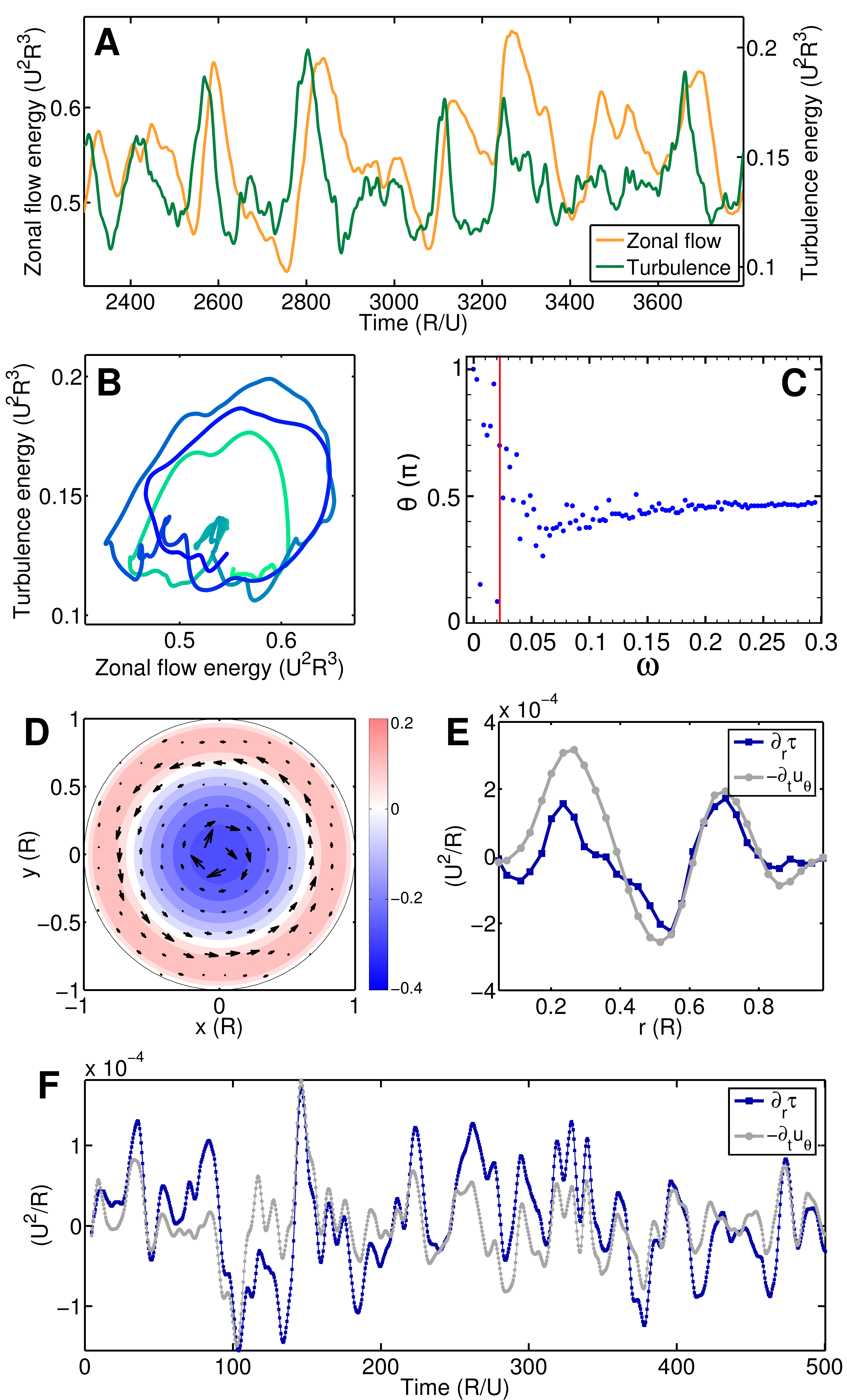}
\caption{\label{fig:1} Predator-prey oscillations in transitional
turbulent pipe flow.  (A) Energy vs. time for the zonal flow (orange)
and turbulent modes (green). (B) Phase portrait of the zonal flow and
turbulent modes as a function of time, with color indicating the
earliest time in dark blue progressing to the latest time in light
green. (C) Phase shift between the turbulent and zonal flow modes as a
function of frequency, showing that the turbulence leads the zonal flow
by $~\pi/2$ consistent with predator-prey dynamics. The phase shift
$\theta\left(\omega\right)= \tan^{-1}\left( Im [\tilde{C}(\omega)] /
Re[\tilde{C}(\omega)]\right)$ and is shifted to be positive, where
$\tilde{C}(\omega)$ is the Fourier transform of the correlation
function between the turbulence and the zonal flow in (A). The red line
corresponds to the dominant frequency in the power spectrum. The phase
shift near small $\omega$ is scatter due to  the finite time duration
of the time series. (D) Velocity field configuration of the zonal flow
mode $\vec{\overline{u}}$. The color bar indicates the value of
$\overline{u}_z$.  (E) Snapshot of the Reynolds stress gradient and
zonal flow time-derivative as functions of $r$. (F) Reynolds stress
gradient and zonal flow time derivative as functions of time. The
evident proportionality shows that zonal flow dynamics is driven by the
radial gradient of the Reynolds stress.}
\end{center}
\end{figure}

\paragraph*{Observation of predator-prey dynamics in Navier-Stokes equations.}

To address these questions without making questionable or
non-systematic approximations,
we have performed direct numerical
simulations (DNS) of the Navier-Stokes equations in a pipe, using the
open-source code \lq\lq Open Pipe Flow" \cite{WK09}, as described in
Appendix \ref{sec:DNS}.
We denote the
time-dependent velocity deviation from the Hagen--Poiseuille flow by
$\vec{u} = (u_z, u_\theta, u_r)$.
Because we were interested in transitional behavior, we looked for a decomposition \cite{prigent2002large,barkley,duguet2013,wesfreid2014} of large-scale modes that would indicate some form of collective behavior, as
well as small-scale modes that would be representative of turbulent dynamics
In particular, we
report here the behavior of the velocity field $(\overline{u}_z,
\overline{u}_\theta, \overline{u}_r)$, where the bar denotes average over $z$ and
$\theta$, and $\overline{u}_r=0$.  We refer to this as the zonal flow (ZF).  In Fourier
space, the zonal flow is given by $\tilde{\vec{u}}(k=0, m=0, r)$, where
$k$ is the axial wavenumber and $m$ is the azimuthal wavenumber, $r$
is the real space radial coordinate and the tilde denotes Fourier
transform in the $\theta$ and $z$ directions only.  Turbulence was
represented by short-wavelength modes, whose energy is $E_T(t)\equiv
\frac{1}{2} \sum_{|k| \geq 1,|m| \geq 1} \int
|\tilde{\vec{u}}(k,m,r)|^2 \,\text{d}V$.

Shown in Figure \ref{fig:1}(A) is a time series for the energy
$E_{ZF}(t)$ of the zonal flow, compared with the energy $E_T(t)$ of the
turbulent energy.  The curves show clear persistent oscillatory
behavior, modulated by long-wavelength stochasticity as shown in the
phase portrait of Figure \ref{fig:1}(B).  In Figure \ref{fig:1}(C), we
have calculated the phase shift between the turbulence and zonal flows,
with the result that the turbulent energy leads the zonal flow energy
by $\sim \pi/2$. This suggests that these oscillations can be
interpreted as a time-series resulting from activator-inhibitor
dynamics, such as occurs in a predator-prey ecosystem.  Predator-prey
ecosystems are characterized by the following behavior: the \lq\lq
prey" mode activates the \lq\lq predator" mode, which then grows in
abundance.  At the same time, the growing predator mode begins to
inhibit the prey mode.  The inhibition of the prey mode starves the
predator mode, and it too becomes inhibited.  The inhibition of the
predator mode allows the prey mode to re-activate, and the population
cycle begins again.

The flow configuration for the predator mode is shown in Figure
\ref{fig:1} (D), and consists of a series of azimuthally symmetric modes
with direction reversals as a function of radius $r$. Such banded shear
flows are known as zonal flows and are of special significance in
plasma physics, astrophysical and geophysical flows, owing to their
role in regulating turbulence \cite{diamond1994}.  The purely
azimuthal component of the zonal flow, denoted by
$\overline{u}_\theta(r)$ is spatially uniform in $z$, and the
lack of a radial component means that it is not driven by pressure
gradients. Thus it can only exist due to nonlinear interactions with
turbulent modes.  In this sense, it is a collective mode, one with
special significance for transitional turbulence.

The simplest way for such an azimuthal shear flow to couple to
turbulent fluctuations is through the Reynolds stress $\tau$: however,
a uniform Reynolds stress cannot drive a shear flow, so the first
symmetry-allowed possibility is the radial gradient of the Reynolds
stress \cite{diamond1994}, as expressed in the
Reynolds momentum equation.  Thus, to probe the dynamics that govern
the emergence of the zonal flow, we have calculated the time-averaged
radial gradient of the instantaneous Reynolds stress, $\tau \equiv
u'_\theta \cdot u'_r$, where $\vec{u'}(z,\theta,r)\equiv \vec{u} -
\vec{\overline{u}}$, and show in
Figure~\ref{fig:1} (F) the 4.5-time-unit-running-mean time series of
$-\partial_t\overline{u}_\theta$ and the radial gradient
$\partial_r\tau$.  Both quantities have been averaged over $0 \leq z
\leq L$, $0 \leq \theta \leq 2\pi$ and $R_0 \leq r < R$, where
$R_0=0.641 R$, and the resulting time series are clearly highly
correlated.

In general, it is the case that zonal flows are driven by statistical
anisotropy in turbulence, but are themselves an isotropizing influence
on the turbulence through their coupling to the Reynolds stress \cite{sivashinsky1985negative,
bardoczi2014experimental,parker2014generation}.  The fact that
turbulence anisotropy activates the zonal flow, and that zonal flow
inhibits the turbulence is responsible for the predator-prey
oscillations observed in the numerical simulations.


\paragraph*{Lifetime of stochastic predator-prey populations.}

Phase transition theory \cite{goldenfeld1992lectures} would suggest
that the emergence of a zonal flow collective mode dominates the
non-equilibrium transition of pipe flow from the laminar to the
turbulent state, through the predator-prey interaction with the small
scale velocity fluctuations.  Such a \lq\lq two fluid" effective field
description of transitional turbulence implies that stochastic
predator-prey populations should undergo spatio-temporal fluctuations
whose functional form matches precisely the observations for the
lifetime and splitting time of turbulent puffs in a pipe.  To test
this idea, we have performed simulations of a spatially-extended stochastic
predator-prey ecosystem, in a quasi-one-dimensional geometry to mimic
the pipe environment.  The specific system has three trophic levels:
nutrient (E), Prey (B) and Predator (A), which correspond in the fluid
system to laminar flow, turbulence and zonal flow respectively.  The
interactions between individual representatives of these levels are
given by the following rate equations
\begin{eqnarray}
&&A_i\xrightarrow{\textit{$d_A$}}E_i,\qquad
B_i\xrightarrow{\textit{$d_B$}}E_i,\qquad
A_i+B_j\xrightarrow[{\langle ij\rangle}]{\textit{$p$}}A_i+A_j,\nonumber\\
&&B_i+E_j\xrightarrow[{\langle ij\rangle}]{\textit{$b$}}B_i+B_j,\qquad
B_i\xrightarrow{\textit{$m$}}A_i,\nonumber\\
&&A_i+E_j\xrightarrow[{\langle ij\rangle}]{\textit{$D$}}E_i+A_j,\;\;\;
B_i+E_j\xrightarrow[{\langle ij\rangle}]{\textit{$D$}}E_i+B_j.
\label{eqn:reactions}
\end{eqnarray}
where $d_A$ and $d_B$ are the death rates of A and B, $p$ is the
predation rate, $b$ is the prey birth rate due to consumption of
nutrient, $\langle ij\rangle$ denotes hopping to nearest neighbor
sites, $D$ is the nearest-neighbor hopping rate, and $m$ is the point
mutation rate from prey to predator, which models the induction of
the zonal flow from the turbulence degrees of freedom.

We are
primarily interested in long-wavelength properties of the system, at
least in the vicinity of the turbulence transition, where we expect the
transverse correlation length to be larger than the pipe diameter,
implying that the behavior is in fact quasi-one-dimensional.  The
crossover phenomena associated with this have been discussed previously
 \cite{sipos2011directed}, and thus our quasi-one-dimensional model
should be appropriate and quantitatively correct near the transition.

\begin{figure}
\begin{centering}
\includegraphics[width=0.8\columnwidth]{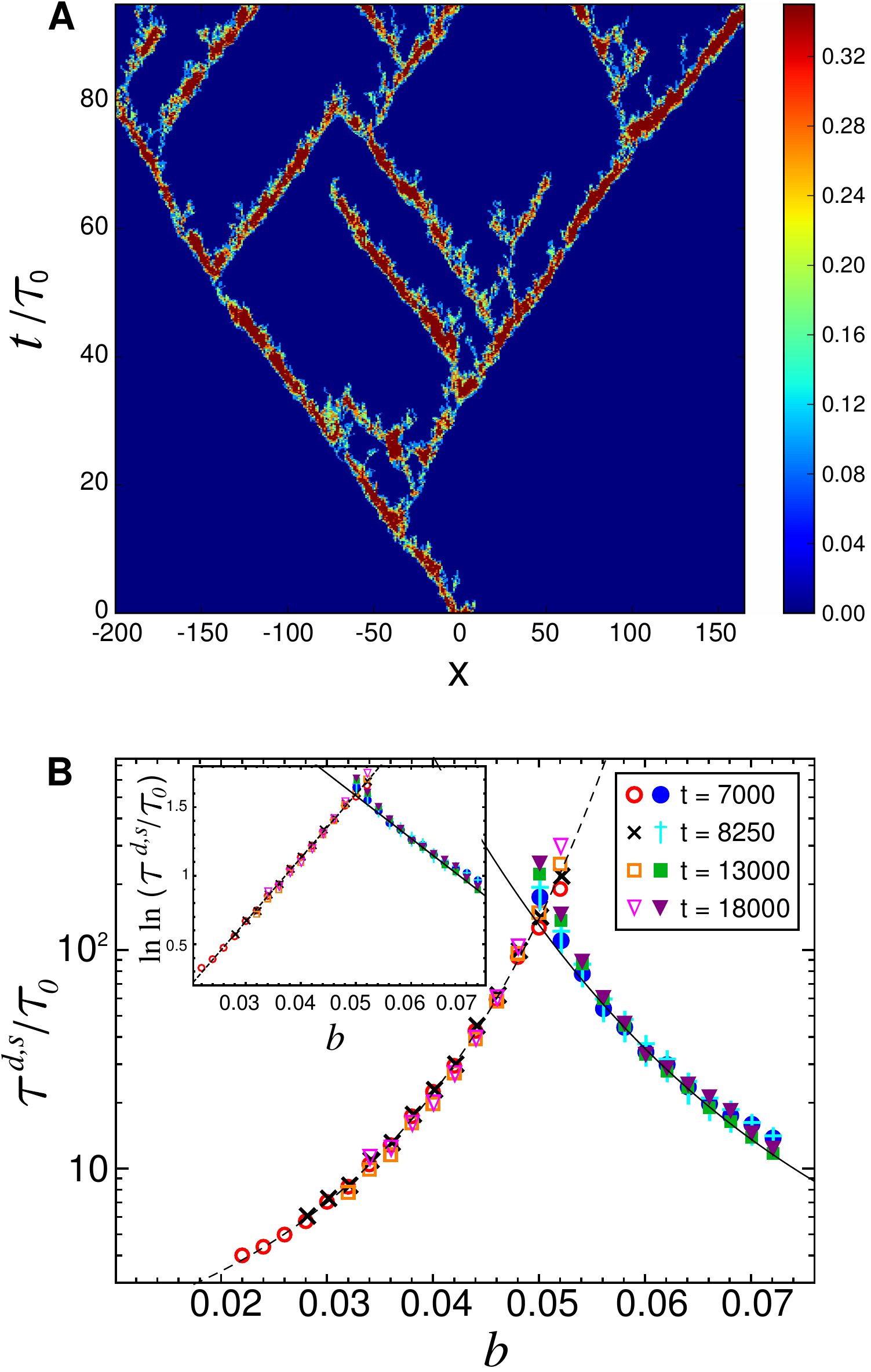}
 \caption{\label{fig:2} Stochastic predator-prey model reproduces the
phenomenology of transitional pipe turbulence.  Decay lifetime,
$\tau^d$, and splitting time, $\tau^s$, of clusters of prey are
memoryless processes and obey super-exponential statistics as a
function of prey birth rate.  To compare with the experiments
\cite{hof_lifetime}, predator-prey dynamics are performed in
two-dimensional pipe geometry as described in the text. (A) World line
of clusters of prey splitting to form predator-prey traveling waves.
The color measures the local density of prey, corresponding to
intensity of turbulence in pipe flow. In the simulation, the
dimensionless parameters are $D=0.1$, $b=0.1$, $p=0.2$, $d_A=0.01$,
$d_B=0.01$ and $m=0.001$. In the model simulated, diffusion is
isotropic, not biased as would be the case corresponding to a mean
flow, where the clusters will accumulate at large times with a
well-defined separation set by the depletion zone of nutrient behind
each predator-prey traveling wave. (B) Log decay lifetime of prey cluster and
splitting time as a function of prey birth rate. The upward curvature
signifies super-exponential behavior. The parameters are $D=0.01$,
$p=0.1$, $d_A=0.015$, $d_B=0.025$ and $m=0.001$. Inset: Double log
lifetime vs prey birth rate, showing the fit to the following
functional forms: the dashed curve is given by $\tau^d/\tau_0 = \exp
(\exp (46.539 b - 0.731))$, and the solid curve is given by
$\tau^s/\tau_0 = \exp (\exp (-31.148 b - 3.141))$.}
\end{centering}
\end{figure}

In our simulation, described in Appendix \ref{sec:StochasticModel}, the control
parameter is the prey birth rate $b$. When $b$ is small enough, the
population is metastable, and cannot sustain itself: all individuals,
both predator and prey, eventually die within a finite lifetime
$\tau^d(b)$.  As $b$ increases, the lifetime of the population increases
rapidly: in particular the prey lifetime increases rapidly with $b$. At
large enough values of $b$, the decay of the initial population is not
observed, but instead the initially localized population proliferates,
spreading outwards and spontaneously splitting into multiple clusters,
as shown in the space-time plot of clusters of prey of Figure
\ref{fig:2} (A).

To quantify these observations, we have measured both the lifetime of
population clusters in the metastable region and their splitting time
using a procedure directly following that of the turbulence experiments
and simulations \cite{avila2011onset}, and described in Appendix
\ref{sec:PredPreySimulation}.  We comment that both timescales involve
implicitly measurements of quantities that exceed a given threshold,
and thus it is natural that the results are found to conform to extreme
value statistics \cite{nigel_evs,sipos2011directed}.

In Figure \ref{fig:2} (A) we show the phenomenology of the dynamics of
initial clusters of prey, corresponding to the predator-prey analogue
for the experiments in pipe flow which followed the dynamics of an
initial puff of turbulence injected into the flow \cite{hof_lifetime}.
Depending upon the prey birth rate, the cluster decays either
homogeneously or by splitting, precisely mimicking the behavior of
turbulent puffs as a function of Reynolds number. The extraction from
data of decay times is described in Appendix \ref{sec:PredPreySimulation}.  In Figure
\ref{fig:2} (B) is shown the semi-log plot of lifetime for both decay
and splitting as a function of prey birth rate, the upward curvature
indicative of super-exponential behavior.  The inset to Figure
\ref{fig:2} (B) shows a double exponential plot of puff lifetime and
splitting time vs. prey birth rate, the straight line being the fit to
the functional form indicated in the caption. These figures indicate a
remarkable similarity to the corresponding plots obtained for
transitional pipe turbulence in both experiments \cite{hof_lifetime}
and direct numerical simulations \cite{avila2011onset}, and demonstrate
conclusively that experimental observations are well captured by an
effective two-fluid model of pipe flow turbulence with predator-prey
interactions between the zonal flow and the small scale turbulence.

\paragraph*{Universality class of the laminar-turbulence transition in pipes.}

\begin{figure}
\begin{centering}
\includegraphics[width=0.95\columnwidth]{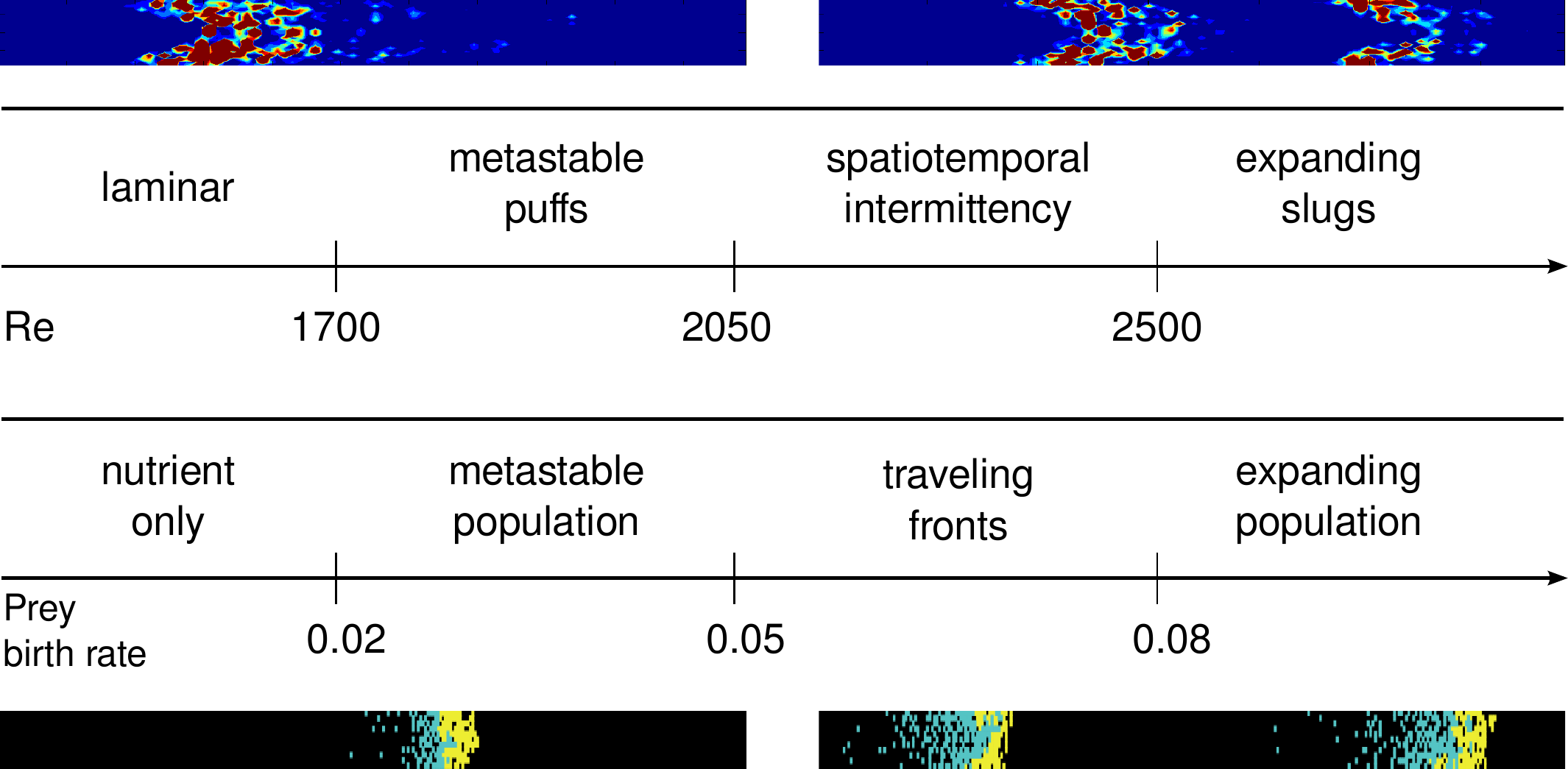}
 \caption{\label{fig:3} Schematic phase diagram for transitional pipe
turbulence as a function of Reynolds number compared with the phase
diagram for predator-prey dynamics as a function of prey birth rate.
Above each phase is shown a typical flow or predator-prey
configuration, indicating the similarity between the turbulent pipe and
ecosystem dynamics.}
\end{centering}
\end{figure}

The two-fluid predator-prey model expressed by Equations
(\ref{eqn:reactions}) exhibits a rich phase diagram that captures the
main features observed in transitional turbulence in pipes.  The
transition to puff-splitting can be identified with a change of
stability of the spatially-uniform mean-field predator-prey coexistence
point, where a stable node becomes a stable focus or spiral with
increasing birth rate.  In the language of predator-prey systems, this
corresponds to the breakdown of spatially homogeneous prey domains into
periodic traveling wave states. The phase diagram is sketched in Figure
\ref{fig:3}, along with the corresponding phase diagram for
transitional pipe turbulence as determined by experiment.  The
phenomenology of the predator-prey system mirrors that of turbulent
pipe flow.

In order to determine the universality class of the non-equilibrium
phase transition from laminar to turbulent flow, we use the two-fluid
predator-prey mode in Equations (\ref{eqn:reactions}).  Near the
transition to prey extinction, the prey population is very small and no
predator can survive, and thus Equations (\ref{eqn:reactions}) simplify to
\begin{eqnarray}
&&B_i\xrightarrow{\textit{$d_B$}}E_i,\qquad B_i+E_j\xrightarrow[{\langle ij\rangle}]{\textit{$b$}}B_i+B_j,\nonumber\\
&&B_i+E_j\xrightarrow[{\langle ij\rangle}]{\textit{$D$}}E_i+B_j.
\label{eqn:DPreactions}
\end{eqnarray}
These equations are exactly those of the reaction-diffusion model for
directed percolation \cite{odor2004universality}. A more detailed and
systematic way to reach this conclusion is to represent Equations
\ref{eqn:reactions} exactly in path integral form using the Doi
formalism
\cite{doi1976,grassberger1980fock,mikhailov1981path,goldenfeld1984kinetics,mattis1998uses,odor2004universality}.
The resulting action simplifies near the transition to that of Reggeon
field theory  \cite{mobilia2007,tauber2012}, which has been shown to be
in the universality class of directed percolation
 \cite{cardy1980,janssen1981}. Numerical simulations of 3 + 1
dimensional directed percolation in a pipe geometry have reproduced the
statistics and behavior of turbulent puffs and slugs in pipe flow
 \cite{sipos2011directed,allhoff2012directed}, and a detailed comparison
between theory and experiment has been presented
 \cite{shi2015universality}.  The super-exponential behavior of DP might
seem to contradict the expectation based upon the known critical
behavior (e.g., see Ref.  \cite{dp_paper}). However, it is important to
recognize that the usual exponents relate to DP starting from a single
seed, whereas the experiments and simulations are conducted with an
extended seed that has a finite length or number of seed points.  These
points behave as independent identically-distributed random variables
as long as the transverse correlation length is much smaller than the
seed size, but once the correlation length is of order the seed size,
the seed is effectively a single correlated extended source, and once
the correlation length is much larger than this size, there will be a
crossover to the usual DP exponents.

\paragraph*{Discussion.}

The observation of the emergence of a zonal flow, excited by the
developing turbulent degrees of freedom and the demonstration of its
role in determining the phenomenology of transitional pipe turbulence
has an interesting consequence: the zonal flow can be assisted by
rotating the pipe, and this should catalyze the transition to
turbulence, causing it to occur at lower Re. Indeed experiments on
axially-rotating pipes \cite{murakami1980turbulent} are consistent with
this prediction.

The idea that predator-prey dynamics can arise in turbulence is by no
means new, and such behavior was proposed by Diamond and collaborators
 \cite{diamond1994,kim2003,itoh2006} many years ago in the context of
the interaction between drift-wave turbulence and zonal flows in
tokomaks; indeed the predator-prey oscillations were recently observed
in tokomaks
 \cite{estrada2010,conway2011,xu2011,estradaPRL2011,schmitz2012} and in
a table-top electroconvection analogue of the L-H transition
 \cite{bardoczi2014experimental}.
The new ingredient we have presented
in this paper is the observation of a zonal flow emerging in the very
simple setting of transitional pipe turbulence, and the demonstration
of predator-prey oscillations that suggest a minimal mesoscale model of
transitional turbulence that accounts for the observed statistical
behavior of puffs.
%
%
%
%
%
The logic of our work is that we used DNS to identify the important
collective modes at the onset of turbulence---the predator-prey
modes---and then wrote down the simplest minimal stochastic model to
account for these observations. This model {\it predicts} without using
the Navier-Stokes equations the puff lifetime and splitting behavior
observed in experiment.  This approach is a precise parallel to that
used in the conventional theory of phase transitions, where one builds
a Landau theory, a coarse-grained (or effective) theory, using symmetry
principles. This intermediate level description can then be used as a
starting point for renormalization group analysis to compute the
critical behavior.

Note that it is neither necessary nor desirable to derive the
coarse-grained effective theory from the microscopic description. Such
a derivation is not necessary, because any such derivation would need a
small parameter and would thus only have limited validity due to the
analytical approximations made. Instead, the effective theory can be
obtained from symmetry principles. A familiar example of this situation
is that even though the Navier-Stokes equations can be derived from
Boltzmann's kinetic equations for gases, such a derivation would imply
that the Navier-Stokes equation is only valid for dilute gases.  In
fact, the Navier-Stokes equations are an excellent description for
dense liquids as well, and can be obtained by perfectly satisfactory
phenomenological and symmetry arguments.  The derivation from
Boltzmann's kinetic theory is inherently limited by the regime of
validity of the kinetic theory---low density---and this leads to an
unnecessarily restrictive derivation of the equations of fluid
dynamics.  The reason why an analytical derivation of the
coarse-grained theory is unnecessary is that even if the coefficients
of the terms could be computed in the order parameter expansion of the
Landau theory, they do not come into the exponents or scaling functions
anyway, and thus they do not affect the critical behavior.  In the
present context, the effective theory we obtain is then mapped exactly
into directed percolation.
%
%
%
%
%
The observation of an emergent zonal flow and predator-prey
oscillations with its attendant minimal \lq\lq two-fluid" model,
provide a direct and unbroken link between the Navier-Stokes equations
and the directed percolation universality class for transitional
turbulence.  Our work underscores not only the potential importance of
zonal flows in other transitional turbulence situations, but also shows
the utility of coarse-grained effective models for non-equilibrium
phase transitions, even to states as perplexing as fluid turbulence.

\emph{Acknowledgements.} We gratefully acknowledge helpful discussions
with Y. Duguet and Z. Goldenfeld.  We especially thank Ashley Willis
for permission to use his code \lq\lq Open Pipe Flow" \cite{WK09}. This
work was partially supported by the National Science Foundation through
grant NSF-DMR-1044901.

\bibliography{transition-turbulence-predator-prey}
\bibliographystyle{apsrev4-1}

\appendix

\begin{center}
{\bf SUPPLEMENTARY MATERIAL}
\end{center}

\renewcommand{\thefigure}{S{\arabic{figure}}}
\setcounter{figure}{0}

\section{Direct numerical simulations of the Navier-Stokes
equations.} \label{sec:DNS} We performed direct numerical simulations
(DNS) of the Navier-Stokes equations in a pipe, using the open-source
code \lq\lq Open Pipe Flow" \cite{WK09}. The equations were solved
using a pseudo-spectral method in cylindrical coordinates \cite{WK09},
having 60 grid points in the radial ($r$) direction, 32 Fourier modes
in the azimuthal ($\theta$) direction and 128 modes in the axial ($z$)
direction. Such a model is of course a reduced description of reality,
but the main features can be well captured, with a slight
renormalization of the $\Re$ needed to compare with experiment
\cite{WK09}.  For the accuracy required in our study, we used 32 modes
in the azimuthal direction, typical of most studies in the literature,
although we note that useful, semi-quantitative results have been
obtained with as few as 2 modes \cite{WK09}. The spatial resolutions
were chosen such that the resolvable power spectra span over six orders
of magnitude. The pipe length $L$ is 20 times its radius $R$, with
periodic boundary conditions in the $z$ direction \cite{WK09}. With
this resolution, the transition to turbulence occurs in a range of
$\Re$ numbers between 2200 and 3000, and moves to smaller $\Re$ at
still higher resolution. We report here measurements at $\Re = 2600$,
slightly above the transition \cite{avila2011onset}. The mass flux and
$\Re = 2600$ were held constant in time \cite{WK09}. The laminar flow
is the Hagen--Poiseuille flow, which was independent of time as the
mass flux was held constant \cite{WK09}.

\begin{figure}[t]
\begin{centering}
\includegraphics[width=0.99\columnwidth]{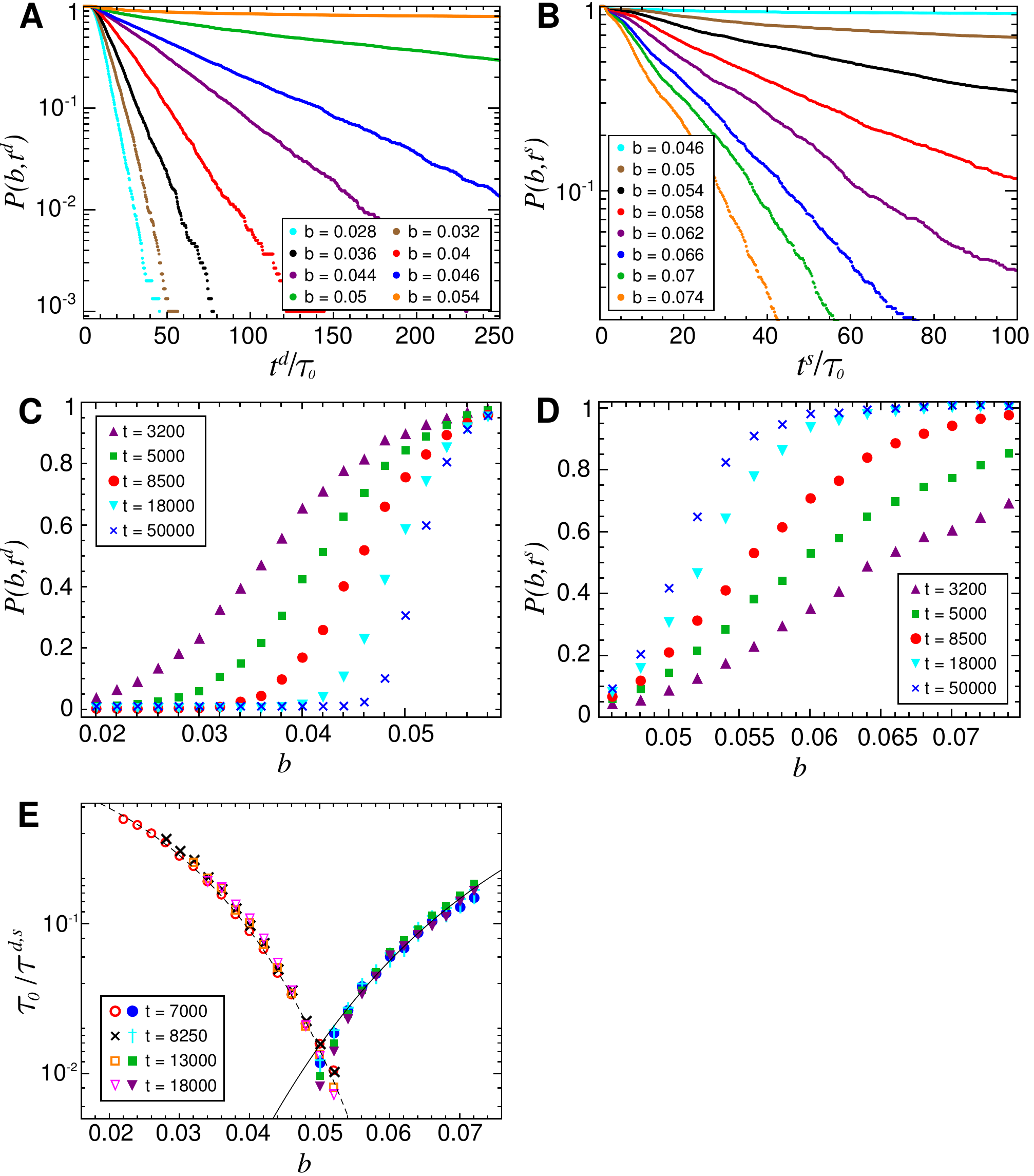}
 \caption{\label{fig:1S} Stochastic predator-prey model reproduces the
phenomenology of transitional pipe turbulence.  Decay lifetime, $\tau^d$, and splitting
time, $\tau^s$, of clusters of prey are memoryless processes and obey
super-exponential statistics as a function of prey birth rate.  To
compare with the experiments \cite{hof_lifetime}, predator-prey
dynamics are performed in two-dimensional pipe geometry as described in
the text. The dimensionless parameters in the simulation are $D=0.01$,
$p=0.1$, $d_A=0.015$, $d_B=0.025$ and $m=0.001$. (A) Log survival
probability of prey cluster vs. time during homogeneous decay to
extinction. Here the characteristic time scale that is estimated by
$\tau_0 \sim 200$.  (B) Log survival probability of prey cluster vs.
time during decay to splitting. (C) Survival probability of prey
cluster as a function of prey birth rate during homogeneous decay to
extinction.  (D) Survival probability of prey cluster as a function of
prey birth rate during decay to splitting. (E) Log inverse lifetime of
prey cluster, as a function of prey birth rate during homogeneous decay
to extinction (left curve, $\tau^d$) and during decay to splitting
(right curve, $\tau^s$). The dashed curve is given by $\tau_0/\tau^d =
1 /\exp (\exp (46.539 b - 0.731))$, and the solid curve is given by
$\tau_0/\tau^s = 1 /\exp (\exp (-31.148 b - 3.141))$. }
\end{centering}
\end{figure}

\section{Stochastic simulations of predator-prey dynamics.}
\label{sec:StochasticModel}

The specific system has three trophic levels: nutrient (E), Prey (B)
and Predator (A), which correspond in the fluid system to laminar flow,
turbulence and zonal flow respectively. Such a system can be naively
modeled by the Lotka-Volterra ordinary differential equations
\cite{lotka1910contribution,volterra1927variazioni,renshaw1993modelling},
which in the case of ecosystems with finite resources do not permit
long-time persistent oscillatory solutions, unless additional
biological details such as functional response are included.  In fact,
it is necessary to include the dynamics of individual birth-death
events, and when this is done correctly, it is found that the number
fluctuations drive the population oscillations
\cite{mckane2005predator} through resonant amplification.  Thus, we use
a stochastic model at the outset.

The interactions between individual representatives of these levels are
given by Equations \ref{eqn:reactions}. We simulated these equations on a $401\times 11$ lattice in two
dimensions, intended to emulate the pipe geometry. Lattice sites were
only allowed to be occupied by one of E, A or B. The predator (A) and
prey (B) are additionally allowed to diffuse via random walk on the
lattice with diffusion coefficient 0.1 in units of the square of the
lattice spacing divided by the time step (set equal to unity). The
initial conditions for the simulations were a random population of prey
and predator, occupying with probability $4/5$ and $1/5$ respectively
on the lattice sites between $x \in [-15,15]$ and $y \in [-5,5]$ where
$x$ labels the direction along the axis of the ecosystem (pipe) and $y$
labels the transverse direction.  The predator-prey dynamics in Eq.
\ref{eqn:reactions} was implemented by the following algorithm: at each
time step, a site $i$ is randomly chosen, a random number $s$, is
generated from the uniform distribution between zero and one. The
behavior on the site is decided by the random number: (1) if $s < 1/6$
and the site $i$ is occupied by any individual, and if a randomly
chosen neighbor site is empty, then that individual diffuses to the
random neighboring site with rate $\mu=0.01$ (i.e. this reaction
happens if another uniformly distributed random number is less than $1
- \exp{\mu}$); (2) if $1/6 \leq s < 1/3$ and the site $i$ is occupied
by a prey individual, and if a randomly chosen neighbor site, $j$, is
empty, then one prey individual is born on the site $j$ with rate $b$;
(3) if $1/3 \leq s < 1/2$ and the site $i$ is occupied by a predator
individual, and if a randomly chosen neighbor site, $j$, is occupied by
a prey individual, then the prey individual is replaced by a new-born
predator individual with rate $p$; (4) if $1/2 \leq s < 2/3$ and the
site $i$ is occupied by a predator individual, that predator individual
dies with rate $d_A$; (5) if $2/3 \leq s < 5/6$ and the site $i$ is
occupied by a prey individual, that prey individual dies with rate
$d_B$; (6) if $5/6 \leq s < 1$ and the site $i$ is occupied by a prey
individual, then the prey individual is replaced by a predator
individual with rate $m$. Then within the same time step, the above
processes are repeated $401 \times 11$ times so that on average
one reaction takes place at each lattice site in the system.\\\\

\section{Measurement of decay and splitting lifetimes.}
\label{sec:PredPreySimulation}

We measured both the lifetime of population clusters in the metastable
region and their splitting time using a procedure directly following
that of the turbulence experiments and simulations
\cite{avila2011onset}. To this end, we monitor the coarse-grained prey
population density
$\tilde{n}_B(i)=\sum_{j=-J}^{j=J}\sum_{l=-H/2}^{l=H/2}n_B(i+j,l)/(H+1)/(2J+1)
- 0.25$, where $H$ is the height of the system (11 lattice units) and
$J=3$. The lifetime of prey clusters is defined as the time it takes
for the last prey individual to die.  The cluster splitting time is
defined as the first time that the distance between the edges of two
coarse-grained prey clusters exceed $25$ unit sites.  We comment that
both timescales involve implicitly measurements of quantities that
exceed a given threshold, and thus it is natural that the results are
found to conform to extreme value statistics
\cite{nigel_evs,sipos2011directed}.


In Figure \ref{fig:1S} we show the phenomenology of the dynamics of
initial clusters of prey, corresponding to the predator-prey analogue
for the experiments in pipe flow which followed the dynamics of an
initial puff of turbulence injected into the flow \cite{hof_lifetime}.
Depending upon the prey birth rate, the cluster decays either
homogeneously or by splitting, precisely mimicking the behavior of
turbulent puffs as a function of Reynolds number. Figure \ref{fig:1S}
(A) and (B) show that the decay is exponential in time, indicating that
it is a memoryless process with a single time constant. Figure
\ref{fig:1S} (C) and (D) show that the survival probability is a
sigmoidal curve, whose inverse lifetime as a function of prey birth
rate is plotted in a log scale in Figure \ref{fig:1S} (E).  If the lifetime were an exponential function, this curve would be
a straight line with negative slope.  The downward curvature is a
manifestation of super-exponential behavior. These figures indicate a
remarkable similarity to the corresponding plots obtained for
transitional pipe turbulence in both experiments \cite{hof_lifetime}
and direct numerical simulations \cite{avila2011onset}, and demonstrate
conclusively that experimental observations are well captured by an
effective two-fluid model of pipe flow turbulence with predator-prey
interactions between the zonal flow and the small scale turbulence.

\end{document}